\documentclass[journal=jpclcd,manuscript=letter]{achemso}

\usepackage[T1]{fontenc}
\usepackage[utf8]{inputenc}
\usepackage{graphicx}
\usepackage{bm}
\usepackage{xcolor}
\usepackage{xr-hyper}
\usepackage{hyperref}
\usepackage{amsmath,amssymb}


\author{Vladislav Korostelev}
\affiliation{Department of Materials Engineering, Auburn University, Auburn, Alabama, USA}
\author{Pjotrs \v{Z}guns}
\affiliation{Institute of Solid State Physics, University of Latvia, Riga, Latvia}
\author{Konstantin Klyukin}
\email{klyukin@auburn.edu}
\affiliation{Department of Materials Engineering, Auburn University, Auburn, Alabama, USA}
\title{Can Universal Machine-Learning Interatomic Potentials Capture Small Polarons? A Geometry-Based Benchmark in Transition-Metal Oxides}
\begin{document}


\begin{tocentry}
\centering
\includegraphics[width=3.25in,height=1.75in]{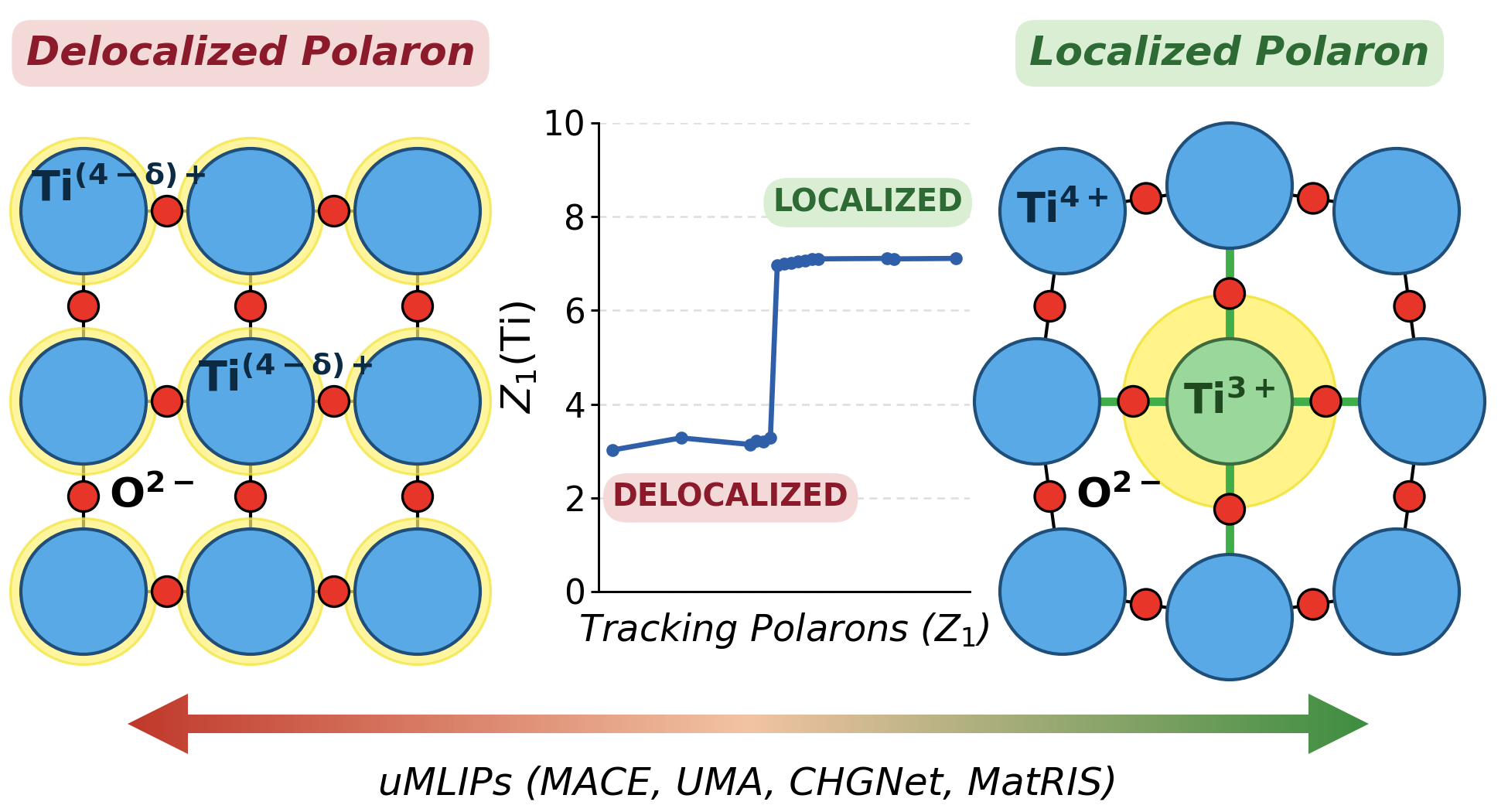}
\end{tocentry}

\begin{abstract}
Small polarons, formed when excess electrons become localized through lattice distortions, play a critical role in charge transport, catalysis, and degradation processes in transition-metal (TM) oxides. Whether universal machine-learning interatomic potentials (uMLIPs), which lack explicit electronic degrees of freedom, can describe them remains unclear. Here, we investigate H-induced electron localization across a range of TM oxides using a geometric descriptor based on TM--O bond-length expansion ($Z$-score) to identify polaronic distortions.  We find that uMLIPs exhibit varying degrees of electron localization, from delocalized states to metastable small polarons, with their behavior strongly dependent on the training data and treatment of electronic correlations. In particular, the absence of Hubbard $U$ corrections for specific TM cations in the training data leads to electron delocalization, in contrast to DFT+$U$ predictions. Although some uMLIPs  reproduce polaron hopping barriers during coupled proton–polaron transfer, all of them fail to stabilize spatially separated polaron–donor configurations, instead favoring localization near the H dopant.
\end{abstract}

\maketitle

Excess charge carriers in transition-metal (TM) oxides often self-trap as small polarons stabilized by lattice distortion around the polaronic center. For an electron polaron, localization reduces the host cation and increases its ionic radius, pushing neighboring oxygen atoms outward~\cite{franchini2021,reticcioli2019,dai2025polaronsreview}. These localized carriers play a central role in the properties and performance of functional oxides, influencing charge transport in LiFePO$_4$ cathodes~\cite{maxisch2006}, photocatalytic  ~\cite{setvin2014,klyukin2017,wu2026tio2,yin2018excess} and photodegradation processes ~\cite{klyukin2018}, and alternations in  materials properties associated with coupled proton--electron incorporation ~\cite{chretien2015,nagatsuka2022,ziwritsch2016,yao2020protonic,wu2023bivo4,Korostelev2026}.

Modeling an excess carrier typically requires an explicit treatment of the electronic degrees of freedom. In practice, polaron studies rely on density functional theory DFT$+U$ or hybrid functionals, which correct the self-interaction error inherent to semi-local DFT and thereby stabilize the localized polaronic state~\cite{deskins2007,yin2018excess}. Localization is then identified by monitoring charge densities and on-site spin moments~\cite{franchini2021,huang2017}: electron trapping on a diamagnetic cation creates a local moment (e.g., $\sim$1\,$\mu_B$ at Ti$^{3+}$ in TiO$_2$)~\cite{deskins2007,morgan2010,elmaslmane2018}, while trapping on a magnetic cation changes the oxidation state and existing moment of the TM center (e.g., Fe$^{3+}\!\to$Fe$^{2+}$ in $\alpha$-Fe$_2$O$_3$)~\cite{adelstein2014}. However, DFT+U and AIMD-based polaron searches are computationally expensive, particularly for the large supercells and long trajectories needed to sample hopping and configurational disorder. Prior works have used purpose-trained machine-learning force fields to accelerate the identification of polaron configurations within DFT itself, complementing conventional occupation-matrix-control approaches~\cite{birschitzky2022mlconfig,birschitzky2024mlprediction,yalcin2026polflow}. Because localization is also accompanied by a characteristic expansion of the surrounding TM--O coordination shell, the local lattice geometry provides an alternative structural signature of polaron formation that can be evaluated without direct electronic-structure information.

Universal machine-learned interatomic potentials (uMLIPs) are trained on millions of DFT/DFT$+U$ calculations and now drive high-throughput screening and long-time molecular dynamics. We benchmark four uMLIPs spanning distinct training-data compositions: CHGNet$_\mathrm{MP}$~\cite{deng2023chgnet}, trained on Materials Project relaxation trajectories (MPtrj); MACE$_\mathrm{MPA\ med}$~\cite{batatia2023mace}, trained on MPtrj combined with Alexandria; MatRIS$_\mathrm{OAM\ med}$~\cite{zhou2026matris}, trained on OMat24, Alexandria, and MPtrj jointly; and UMA$_\mathrm{OMAT\ s1p2}$~\cite{wood2025uma}, trained on OMat24. In parallel, polaron-specific architectures have been developed that explicitly encode charge states, allowing localized carriers to be tracked during finite-temperature dynamics and extending polaron-hopping simulations to nanosecond timescales~\cite{birschitzky2025leopold,leoni2026surfacepolarons,ahart2026deeppolaron}. Recently, the PolFlow workflow adapted VASP on-the-fly force fields to polarons by labeling the trapping atoms as a separate species. This enables automated, site-selective polaron localization in TiO$_2$(110) at nearly 50\% lower DFT cost~\cite{yalcin2026polflow}. In contrast to these purpose-built architectures, which are typically trained on relatively small, system-specific datasets, it remains unclear whether general-purpose uMLIPs, trained without an explicit representation of oxidation state or localized charge, can recover polaron localization from energies and forces alone. The localized and delocalized solutions are distinct electronic states, differing in spin density, in-gap levels, and metal-insulator character, that map onto subtly different geometries.  uMLIPs trained only on energies and forces have no explicit knowledge of these electronic distinctions, leaving open the question of whether they can nevertheless reproduce the subtle symmetry-breaking lattice distortion associated with a localized polaron: an expanded TM–O coordination shell embedded within an otherwise uniform sublattice.

Several uMLIPs additionally predict site-resolved magnetic moments: CHGNet~\cite{deng2023chgnet}
and MatRIS~\cite{zhou2026matris} are trained on scalar moment magnitudes
alongside energies and forces. Predicted magnetic moment could serve as an independent indicator of polaron localization, complementing a geometric descriptor. Prior CHGNet-based study have modeled multivalent charge ordering, including Mn$^{3+}$ disproportionation in Li$_{0.5}$MnO$_2$ and V$^{3+}$/V$^{4+}$ ordering in Na$_2$V$_2$(PO$_4$)$_3$~\cite{deng2023chgnet,zhong2025mnrich}. In these systems, approximately one charge carrier is introduced for every two transition-metal sites, a concentration sufficiently high to promote collective, extended charge ordering rather than isolated-carrier localization. By contrast, the performance of uMLIPs in the dilute-carrier regime, in which a supercell contains a single excess electron that must localize on one among several chemically equivalent transition-metal sites, has received considerably less attention. This regime is particularly relevant to isolated small polarons generated by dilute doping, intrinsic defects, or charge injection, where localization is governed by subtle local electronic and structural distortions. Although Materials Project, Alexandria, and OMat24 span numerous transition-metal oxidation states, they do not explicitly account for such point-defect configurations, and training data for isolated small polarons is essentially absent. 

\begin{figure*}[ht]
\includegraphics[width=\textwidth]{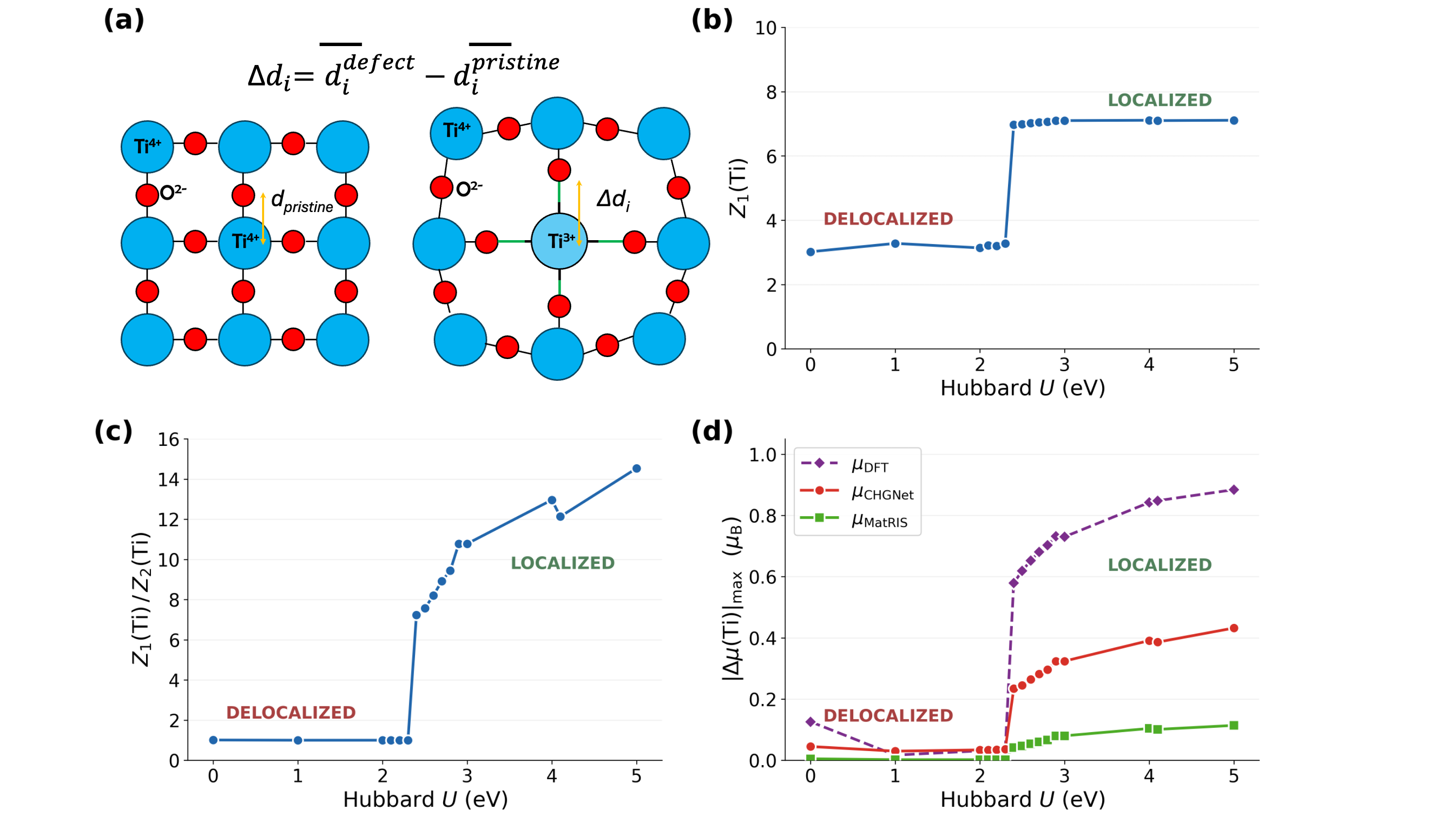}
\caption{Validation of the geometric descriptor on H-doped rutile TiO$_2$ relaxed with DFT$+U$ as a function of $U$(Ti). (a)~Schematic of polaron formation: reduction of Ti$^{4+}$ to Ti$^{3+}$ elongates the surrounding Ti--O bonds by $\Delta d_i = d_i^{\mathrm{defect}} - d_i^{\mathrm{pristine}}$, which is the geometric signature exploited by the $Z$-score descriptor.
(b)~Maximum $Z$-score $Z_1$ and (c)~$Z_1/Z_2$ jump discontinuously at the localization transition, $U \approx 2.3$~eV.
(d)~Maximum on-site moment change $|\Delta\mu(\mathrm{Ti})|_{\max}$ from DFT$+U$ (purple) and from CHGNet (red) and MatRIS (green) single points on the same geometries.}
\label{fig:metric}
\end{figure*}

In this Letter, we systematically assess whether existing uMLIPs can capture polaron localization. We first (i) introduce a purely geometric descriptor that requires no magnetic moments or charge analysis to monitor electron localization, (ii) validate it against Hubbard-$U$ DFT calculations, and (iii) apply it to classify polaron localization on the potential-energy surfaces (PES) of five uMLIPs across several oxides. Presently, uMLIPs take only atomic positions and element species as input and cannot specify excess electron directly. We therefore introduce the excess electron chemically by doping oxides with a neutral interstitial H or Li atom (H$\to$H$^+ + e^-$, Li$\to$Li$^+ + e^-$), or by introducing an oxygen vacancy. The donated electron can either localize as a small polaron on one TM site or delocalize over the TM sublattice.    Computational details, including the DFT$+U$ setup, supercells, and uMLIPs versions, are given in the Supporting Information (SI).

To quantify polaron localization, we introduce a geometric descriptor based on the local oxygen distortion. As shown in Fig.~\ref{fig:metric} (a), polaron formation expands the first oxygen coordination shell around the host TM site, consistent with earlier DFT$+U$ studies establishing this Ti--O bond elongation as the structural signature of a localized Ti$^{3+}$ polaron in TiO$_2$~\cite{deskins2007,ahart2026deeppolaron}.  For each TM site $i$ we compute the mean bond length to its first oxygen shell, $\bar d_i$, in the defective and pristine supercells, and define $\Delta d_i = \bar d_i^{\,\mathrm{def}} - \bar d_i^{\,\mathrm{prist}}$. 
We then convert $\Delta d_i$ into per-site $Z$-scores over the TM sublattice by subtracting the sublattice mean and dividing by the corresponding standard deviation:
\begin{equation}
Z_i = \frac{\Delta d_i - \mu}{\sigma},
\end{equation}
where $\mu$ and $\sigma$ are the mean and standard deviation of $\Delta d_i$ within the corresponding TM sublattice. For mixed-CN lattices, these quantities are calculated separately for each coordination number (CN). After sorting the scores in descending order, $Z_1 \ge Z_2 \ge \dots$, the largest  $Z_1$ value measures how strongly the most-distorted site has expanded, while the $Z_1/Z_2$ quantifies whether the distortion is concentrated on a single TM site or distributed over multiple sites. A localized small polaron is identified by $Z_1/Z_2 > 1$, whereas $Z_1/Z_2 \approx 1$, and similarly $Z_1/Z_3 \approx 1$, indicates a delocalized distortion with no uniquely selected TM site. For bipolarons, as in V$_2$O$_5$, where two TM sites distort simultaneously, we use the analogous metric $(Z_1{+}Z_2)/2Z_3$. The first coordination sphere cutoffs used for each oxide are listed in the SI.

\begin{figure*}[ht]
\includegraphics[width=\textwidth]{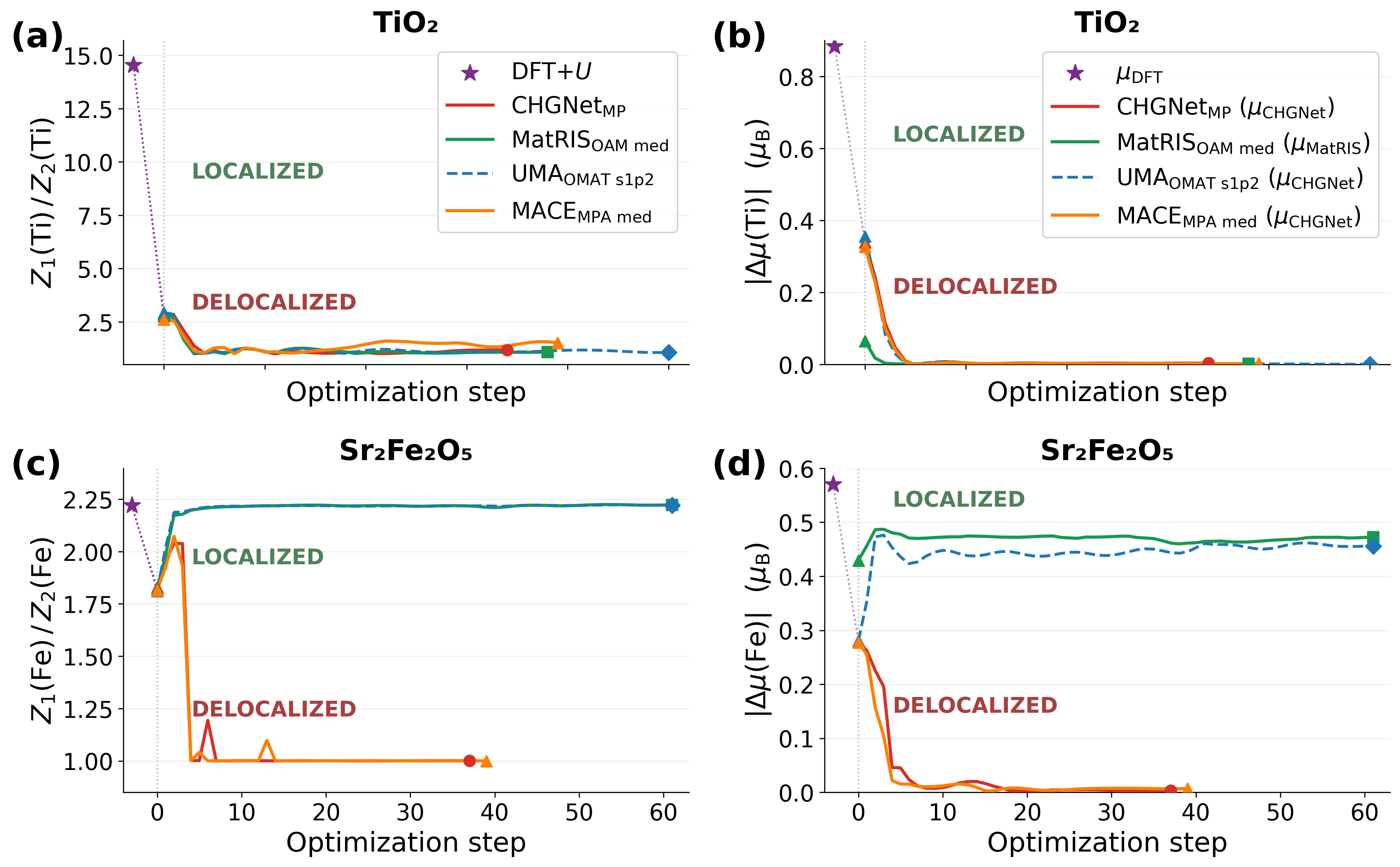}
\caption{uMLIP relaxation of DFT$+U$-localized polaron geometries.
$Z_1/Z_2$ (a,c) and magnetic-moment change $|\Delta\mu|$ (b,d) along BFGS
trajectories initialized from the DFT-localized geometry, for TiO$_2$ (a,b)
and Sr$_2$Fe$_2$O$_5$ (c,d). Stars: DFT$+U$ reference.
$|\Delta\mu|$ is evaluated with each model's own magnetic moments for
CHGNet$_\mathrm{MP}$ and MatRIS$_\mathrm{OAM\ med}$; UMA$_\mathrm{OMAT\ s1p2}$
and MACE$_\mathrm{MPA\ med}$ do not output magnetic moments natively, so their
relaxed geometries are re-evaluated with CHGNet to obtain $\mu_\mathrm{CHGNet}$.
All models delocalize the TiO$_2$ polaron; UMA and MatRIS preserve the
Fe polaron in Sr$_2$Fe$_2$O$_5$, while CHGNet and MACE-MPA do not.}
\label{fig:traj}
\end{figure*}

We first validate the proposed descriptor using rutile TiO$_2$ with one interstitial H, relaxed with DFT$+U$ across a range of  $U$(Ti) values.  As shown in  Figure~\ref{fig:metric},  the localization transition appears at $U \approx 2.3$~eV, as evidenced by change in both the $Z_1$ (see Fig.~\ref{fig:metric}(b)) and $Z_1/Z_2$ (see Fig.~\ref{fig:metric}(c)) metrics, the latter increasing abruptly from $\sim$1 to $>7$, 
clearly separating the small polaron localized on a single Ti site for U => 2.3 eV from the delocalized solution for U <2.3, consistent with the loss of localization reported for DFT$+U$ below $U\approx 2$~eV~\cite{yin2018excess,finazzi2008}. At larger $U$, the continued increase in $Z_1/Z_2$ indicates progressively stronger localization. Magnetic moments independently confirm the same transition: $|\Delta\mu(\mathrm{Ti})|_{\max}$ from DFT$+U$ rises sharply at the same $U \approx 2.3$~eV, reaching $\sim$0.88~$\mu_B$ (Ti$^{4+}\!\to$Ti$^{3+}$) at $U=5$~eV, mirroring the jump in $Z_1/Z_2$ (see Fig.~\ref{fig:metric}(d)).

To test whether selected uMLIPs can recover the magnetic signature of a localized polaron when provided with the distorted geometries, we evaluated CHGNet and MatRIS single-point magnetic moments on the same DFT$+U$
structures. As seen in Fig.~\ref{fig:metric}(c),  CHGNet qualitatively captures the localized polaronic state but substantially underestimates the moment, reaching only $\sim 0.43~\mu_B$. In contrast, MatRIS moments remain considerably smaller of $\sim 0.1~\mu_B$, across the same structures. These results suggest that the CHGNet and MatRIS training sets may not sample Ti magnetic configurations broadly enough to reproduce the DFT$+U$ moment, even for geometries that already contain the structural distortions associated with polaron localization. The geometric $Z$-score therefore provides a more transferable measure of polaron localization that can be easily adopted by any uMLIPs architecture regardless of built-in magnetic moment predictors.

Next we test  whether the PES of a uMLIP supports a localized polaronic minimum. We make this test deliberately favorable to the uMLIPs by initializing each relaxation from the
DFT$+U$ geometry with the polaronic distortion. If a localized minimum exists on the uMLIP PES near this configuration, the relaxation should remain in the polaronic basin. Conversely, delocalization from this starting point indicates that the uMLIP PES does not support a localized minimum in the vicinity of the DFT$+U$ solution.  Initialization from a delocalized geometry is compared in Fig.~S1. We perform this test for all five models and track both $Z_1/Z_2$ and $|\Delta\mu|$ from CHGNet single-point evaluations along the BFGS optimization trajectory (see Fig.~\ref{fig:traj}, relaxation details in the SI).

These results  reveal a systematic limitation of current uMLIPs in describing Ti-polaron localization in rutile TiO$_2$. Despite being initialized from the DFT$+U$ localized geometry, the polaronic distortion collapses within the first few optimization steps for all four tested uMLIP models. This is reflected by the rapid drop of $Z_1/Z_2$ from the DFT$+U$ value of $\sim 14.5$ to $\sim 1$ (Fig.~\ref{fig:traj}(a)), accompanied by the disappearance of the CHGNet-evaluated magnetic moment (Fig.~\ref{fig:traj}(b)). Thus the models fail to preserve the localized polaron state. The origin of this behavior is consistent with the $U$-dependent DFT analysis in Fig.~\ref{fig:metric}: the Ti polaron becomes stable only above $U \approx 2.3$~eV. Because the Materials Project Hubbard correction scheme  used in the underlying training data applies no Hubbard $U$ to Ti, TiO$_2$ configurations in the training set lie on the delocalized side of this threshold.

For Sr$_2$Fe$_2$O$_5$, where the training data use $U_{\mathrm{Fe}} = 5.3$~eV, 
polaron localization is possible in uMLIPs (Fig.~\ref{fig:traj}(c,d)). However, the models behave differently.
UMA and MatRIS preserve the small polaron on the Fe site: $Z_1/Z_2$ remains 
close to the DFT$+U$ value, and CHGNet-evaluated magnetic-moment 
changes reach $\sim 0.45$--0.48~$\mu_B$, consistent with the DFT$+U$ 
Fe$^{3+}\!\to$Fe$^{2+}$ reduction. In contrast, CHGNet and MACE-MPA delocalize the polaron within $\sim$5 
optimization steps. These results show that a sufficiently large training $U$ is necessary but not 
sufficient for polaron stabilization. Additional factors, such as the 
training set's coverage of off-equilibrium (i.e., distorted, non-minimum-energy) configurations and defect content, are discussed below.

We further demonstrate that UMA and MatRIS can each resolve two distinct minima on the PES for Sr$_2$Fe$_2$O$_5$ and V$_2$O$_5$: one with the polaron localized, one delocalized. As seen from Figs.~S2--S3, the two models do not always agree on which minimum is energetically lower or how large an energy difference separates them. For Sr$_2$Fe$_2$O$_5$,  UMA places the localized state $+297$~meV above the delocalized minimum, separated by a barrier of only $\sim$88.6~meV. MatRIS finds a smaller gap ($+36$~meV) but a substantially larger barrier ($156$~meV), agreeing with UMA that the delocalized state is lower in energy. Notably, DFT$+U$ calculations with $U=5.30$~eV strongly favor electron localization and do not stabilize a corresponding delocalized state. For V$_2$O$_5$, DFT$+U$ places the localized single-polaron state $118.5$~meV below the delocalized state, favoring localization. MatRIS reproduces this ordering, placing the localized state $45$~meV below the delocalized minimum behind a $113$~meV barrier, whereas UMA reverses it, placing the localized state $+178$~meV above the delocalized minimum with an $\sim$3.4~meV barrier. Both models therefore capture a genuine transition between the two states, but the relative stability of the localized and delocalized minima remains sensitive to training data and model architecture.

\begin{figure*}[!htbp]
\includegraphics[width=0.6\columnwidth]{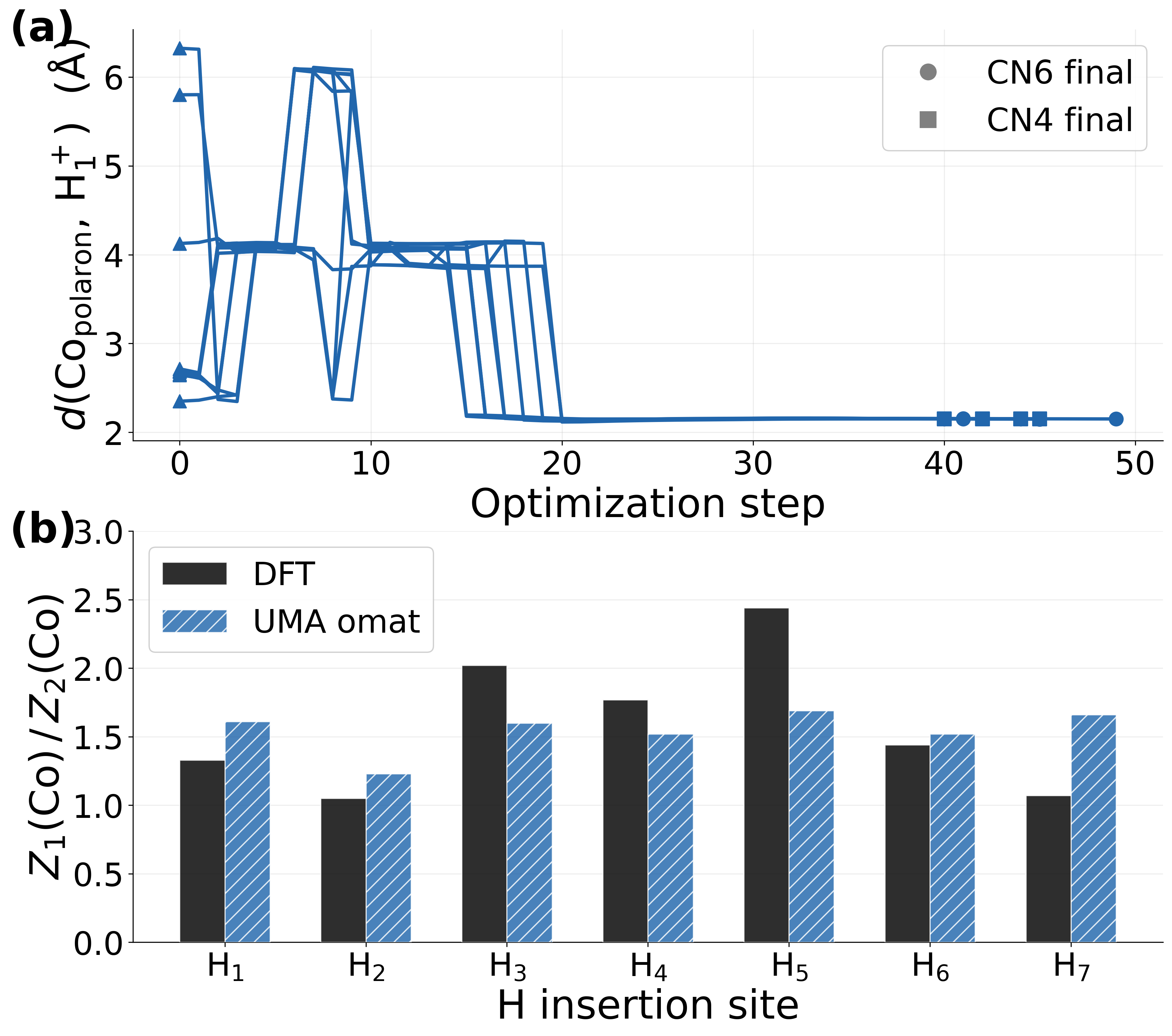}
\caption{Polaron site selectivity in Sr$_2$Co$_2$O$_5$ (UMA).
(a)~Distance between the polaronic Co site and the fixed proton (H$_1$) during relaxations initialized with the distortion on 11 different Co sites; all converge to one site at $\sim$2.15~\AA.
(b)~$Z_1/Z_2$ per CN sublattice for proton sites H$_1$--H$_7$, DFT (black) vs UMA (blue).}
\label{fig:sco}
\end{figure*}

Even uMLIPs that stabilize a localized polaron do not reproduce the spatially separated polaron--proton pairs accessible at the DFT$+U$ level. As shown in our previous work, DFT$+U$ can localize the polaron on different TM sites at a fixed proton position, yielding a family of such pairs whose energies span $\sim 0.05$--0.5~eV, largely reflecting the electrostatic potential energy of defect separation~\cite{Korostelev2026}. On the considered uMLIP PES, however, these separated pairs are not metastable: regardless of the initial polaron site, relaxation collapses the pair onto a single preferred TM site adjacent to the proton. Figure~\ref{fig:sco} illustrates this collapse for brownmillerite Sr$_2$Co$_2$O$_5$. For a proton fixed at the H$_1$ site, we performed 11 independent UMA relaxations, each initialized with the polaronic distortion on a different Co site $\sim$2--6~\AA{} from the proton, spanning both tetrahedral (CN4) and octahedral (CN6) sublattices. Every relaxation converges to the same Co site, $\sim$2.15~\AA{} from the proton (Fig.~\ref{fig:sco}(a)). Repeating the analysis for seven distinct proton positions, H$_1$--H$_7$, we find that UMA stabilizes a localized polaron in every case, with $Z_1/Z_2 = 1.2$--$1.7$, comparable to the DFT$+U$ degree of localization (Fig.~\ref{fig:sco}(b)). It is important to note that the identity of the preferred localization sites at Sr$_2$Co$_2$O$_5$ does not always match the DFT$+U$ ground state calculations. We attribute these discrepancies to near-degenerate neighboring Co sites, with energy differences of tens of meV that the current uMLIPs do not fully resolve. The Fe-based brownmillerite Sr$_2$Fe$_2$O$_5$ shows analogous behavior (Fig.~S5).

\begin{figure*}[!htbp]
\includegraphics[width=0.6\columnwidth]{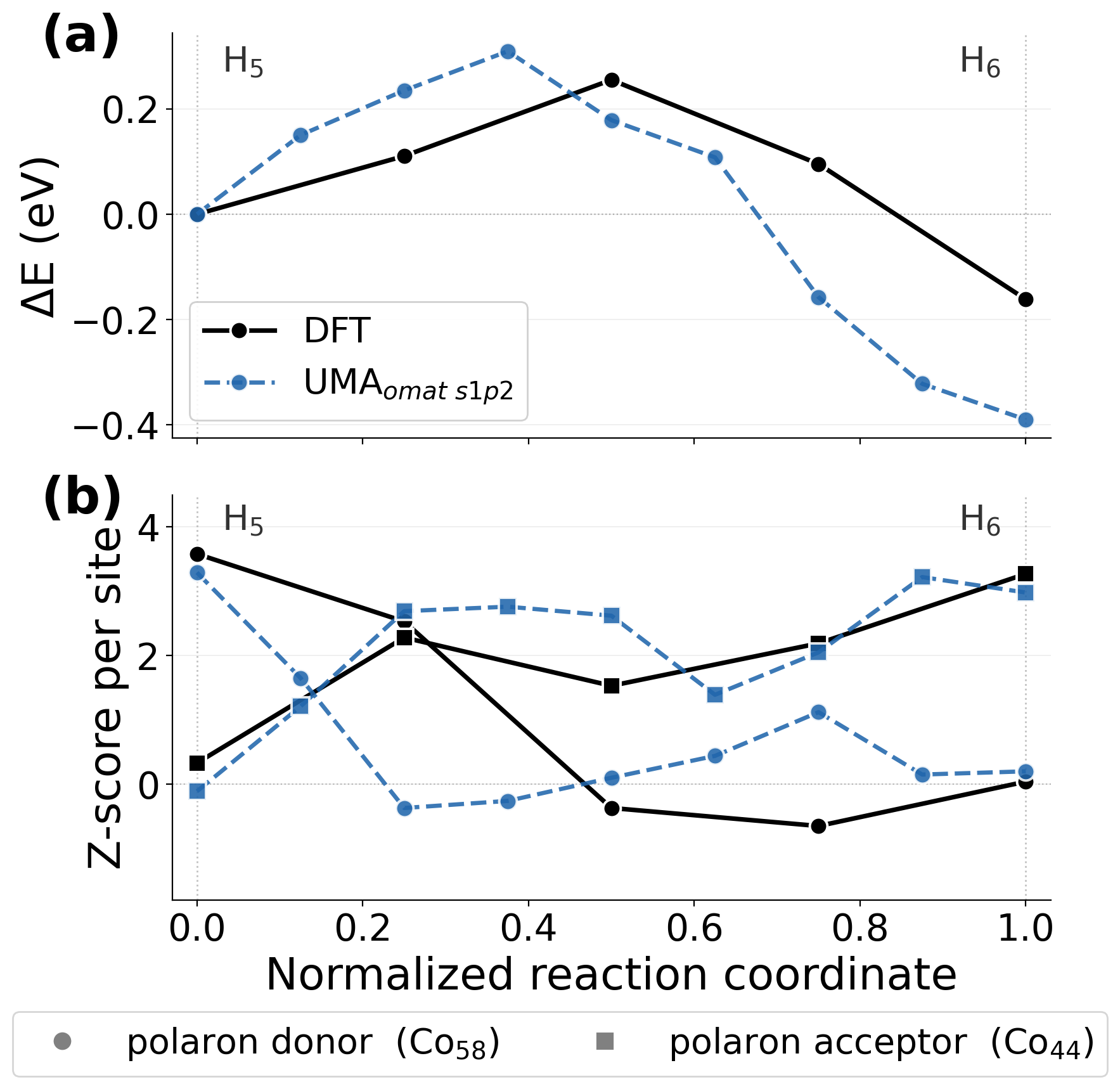}
\caption{H$_5$$\to$H$_6$ polaron transfer in Sr$_2$Co$_2$O$_5$.
(a)~Energy profile: DFT$+U$ CI-NEB (black) vs UMA (blue).
(b)~$Z$-scores of donor (Co\#58, circles) and acceptor (Co\#44, squares) sites along the normalized reaction coordinate, showing the geometric handoff.}
\label{fig:neb}
\end{figure*}

\begin{figure*}[!htbp]
\includegraphics[width=\textwidth]{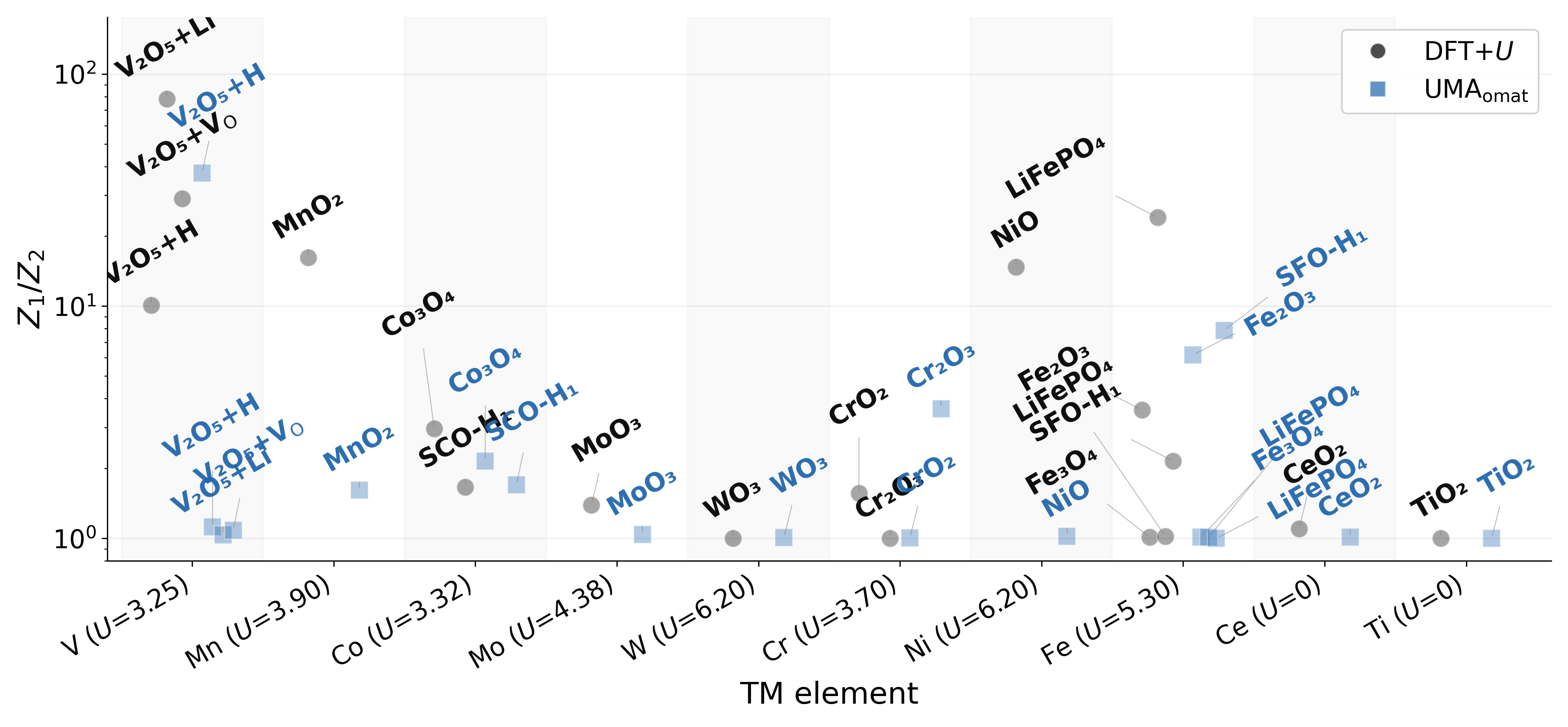}
\caption{$Z_1/Z_2$ per TM element (ordered by Materials Project $U$), DFT$+U$ (black) vs UMA (blue), for H-inserted oxides. For V$_2$O$_5$, Li insertion and oxygen-vacancy configurations are also included.}
\label{fig:trends}
\end{figure*}

Nevertheless, both UMA and MatRIS can reproduce coupled proton--polaron transfer barriers. Figure~\ref{fig:neb} compares the H$_5$$\to$H$_6$ transfer path in Sr$_2$Co$_2$O$_5$, in which the proton rotates from one oxygen plane to another while the polaron hops correspondingly. UMA reproduces the barrier magnitude to within $\sim$50~meV, but overstabilizes the product state ($-0.39$ vs.\ $-0.16$~eV). 
The $Z$-score tracks the polaron hopping from donor to acceptor geometrically (Fig.~\ref{fig:neb}(b)). The donor Co\#58 $Z_1$ decreases from $\sim$3.5 to $\sim$0 while the acceptor Co\#44 rises from $\sim$0 to $\sim$3.5 across the reaction coordinate, confirming that UMA captures the continuous donor-to-acceptor polaron transfer rather than only the initial and final states. MatRIS shows a similar donor--acceptor $Z$-score crossover and a comparable barrier profile along the same reaction coordinate (Fig.~S6).

Finally, we extend the benchmark to a broader set of oxides to determine whether the ability to preserve localization follows the electronic-structure treatment represented in its training data. As in Fig.~\ref{fig:traj}, UMA relaxations for each oxide are initialized from the DFT$+U$-localized polaron geometry. Figure~\ref{fig:trends} compares the DFT$+U$ and UMA values of $Z_1/Z_2$ for oxides containing nine TM elements, with the element-specific Hubbard $U$ values adopted by the Materials Project indicated for reference~\cite{jain2013}.  The comparison reveals broad agreement in the classification of localized and delocalized states. Strongly localizing systems are generally reproduced well, with Co, Fe, V, and Cr oxides exhibiting clearly identifiable polaronic distortions in both DFT$+U$ and UMA. Although the magnitude of $Z_1/Z_2$   varies considerably among these hosts, UMA-based simulations demonstate the ability toreproduce a distinct polaronic site. The applied $+U$ alone, however, does not determine localization. For example, although UMA localizes the excess electron introduced by interstitial H, electrons donated by interstitial Li or an O vacancy remain delocalized in V$_2$O$_5$ (Fig.~S4). Similarly, Cr$_2$O$_3$ localizes the excess carrier, whereas CrO$_2$ remains delocalized in both DFT$+U$ and UMA calculations, demonstrating that the host structure and local coordination environment remain important even for the same TM element. By contrast, NiO, WO$_3$, and CeO$_2$ show no distinct polaronic site in either DFT$+U$ or UMA ($Z_1/Z_2 \approx 1$). Localization in these materials generally requires a larger, compound-specific $U$ or a hybrid functional~\cite{zguns2026,castleton2019}. Consequently, localized configurations are not represented at the electronic-structure level from which the relevant training data were generated.

In summary, we have shown that selected universal machine-learning interatomic potentials can capture polaronic defects in transition-metal oxides, but only to the extent that localized states are represented in their training data, through both the reference theory and the diversity of structural configurations included in training.  To track this behavior, we introduce a geometry-based metric for identifying small polaron localization in transition-metal oxides using only atomic coordinates. The metric quantifies the expansion of the first TM–O coordination shell through site-resolved $Z$-scores, while the $Z_1/Z_2$  distinguishes the distorted polaronic site within the same TM sublattice. Because it requires neither charge densities nor magnetic moments, the metric can be applied directly to machine learning interatomic potentials that do not explicitly describe electronic degrees of freedom. Across the oxides and uMLIPs examined here, it reliably differentiates localized and delocalized states and tracks the transfer of a polaron between neighboring sites.

Applying this metric reveals a strong connection between polaron localization and the electronic-structure level represented in the training data. All tested models delocalize polarons on Ti and Ce, as well as in several other hosts for which the reference calculations do not stabilize a localized state. By contrast, UMA and MatRIS preserve localization in several Co-, Fe-, V-, and Mn-based oxides for which the reference DFT calculations favor localized solutions. The applied Hubbard U value, however, is not by itself sufficient to determine localization: the host structure, local coordination, defect chemistry, and carrier source also influence whether a distinct polaronic minimum is recovered.  Our benchmarks expose important limitations even in localization-capable models. UMA and MatRIS support both localized and delocalized minima in Sr$_2$Fe$_2$O$_5$ and V$_2$O$_5$, but they disagree on their relative stability and delocalization barriers. Moreover, these models collapse the range of spatially separated polaron–donor configurations accessible in DFT+$+U$  onto a single preferred TM site, typically the site nearest the dopant. They therefore reproduce the existence of a local distortion without fully resolving the competing localized states that define the polaron energy landscape. Local transport can nonetheless be captured: UMA reproduces the coupled proton–polaron transfer barrier within approximately 50 meV, and the site-resolved $Z$-scores directly resolve the geometric handoff between donor and acceptor sites. 

Improving the description of polaron energetics requires moving beyond the structural diversity of existing training datasets to explicitly account for localized electronic states.  The limitations of models trained on the OMat24 and Alexandria GGA datasets illustrate that structural diversity and off-equilibrium sampling alone are insufficient, partly because the underlying GGA reference calculations often fail to stabilize localized polaronic states~\cite{barroso2024omat24,schmidt2024alexandria}. Emerging meta-GGA datasets, including MatPES~\cite{kaplan2025matpes} and MP-ALOE~\cite{mpaloe2025}, offer promising opportunities, although accurate localization also requires explicit sampling of symmetry-broken configurations. Incorporating competing localized states and their relative energetics from DFT+$U$, spin-polarized meta-GGA, or hybrid-functional calculations could substantially improve model reliability.

\begin{acknowledgement}
This work was supported in part by the National Science Foundation through DMREF-2324156.  V.~K. acknowledges financial support from the Alabama Graduate Research Scholars Program (GRSP), funded through the Alabama Commission on Higher Education and administered by the Alabama EPSCoR. This work was completed in part with resources provided by the Auburn University Easley Cluster.
\end{acknowledgement}
\appendix

\begin{suppinfo}
Computational details (VASP settings, supercell compositions and magnetic orderings, uMLIP checkpoints, $Z$-score shell cutoffs, workflow) and Figures S1--S6 (geometry dependence, interpolated potential-energy surfaces, V$_2$O$_5$ Li/O-vacancy donors, Sr$_2$Fe$_2$O$_5$ site selectivity, MatRIS proton--polaron transfer). DFT$+U$ geometries and figure data are available at \url{https://doi.org/10.6084/m9.figshare.33758275}.
\end{suppinfo}

\bibliography{references}   

@article{franchini2021,
  author  = {Franchini, Cesare and Reticcioli, Michele and Setvin, Martin and Diebold, Ulrike},
  title   = {Polarons in materials},
  journal = {Nat. Rev. Mater.},
  volume  = {6},
  pages   = {560},
  year    = {2021}
}

@article{finazzi2008,
    author  = {Finazzi, Emanuele and Di Valentin, Cristiana and Pacchioni, Gianfranco and Selloni, Annabella},
    title   = {Excess Electron States in Reduced Bulk Anatase {TiO$_2$}: Comparison of Standard {GGA}, {GGA}+{U}, and Hybrid {DFT} Calculations},
    journal = {J. Chem. Phys.},
    volume  = {129},
    pages   = {154113},
    year    = {2008},
    doi     = {10.1063/1.2996362}
  }

@article{yin2018excess,
  author  = {Yin, Wen-Jin and Wen, Bo and Zhou, Chuanyao and Selloni, Annabella and Liu, Li-Min},
  title   = {Excess Electrons in Reduced Rutile and Anatase {TiO$_2$}},
  journal = {Surf. Sci. Rep.},
  volume  = {73},
  number  = {2},
  pages   = {58--82},
  year    = {2018},
  doi     = {10.1016/j.surfrep.2018.02.003}
}

@unpublished{ahart2026deeppolaron,
  author = {Ahart, Christian S. and Li, Denan and Blumberger, Jochen and Liu, Shi},
  title  = {Polaron Transport in {TiO$_2$} from Machine Learning Molecular Dynamics},
  year   = {2026},
  note   = {arXiv:2606.01763}
}

@article{klyukin2017,
    author = {Klyukin, Konstantin and Alexandrov, Vitaly},
    title = {CO2 Adsorption and Reactivity on Rutile
TiO2(110) in Water: An Ab Initio Molecular
Dynamics Study},
    journal = {The Journal of Physical Chemistry C},
    volume = {121},
    number = {19},
    pages = {10476-10483},
    year = {2017},
    month = {05},
    issn = {1932-7447},
    doi = {10.1021/acs.jpcc.7b02777},
    url = {https://doi.org/10.1021/acs.jpcc.7b02777},
    eprint = {https://pubs.acs.org/jpccck/article-pdf/121/19/10476/7782236/jp7b02777.pdf},
}

@article{yao2020protonic,
  title={Protonic solid-state electrochemical synapse for physical neural networks},
  author={Yao, Xiahui and Klyukin, Konstantin and Lu, Wenjie and Onen, Murat and Ryu, Seungchan and Kim, Dongha and Emond, Nicolas and Waluyo, Iradwikanari and Hunt, Adrian and Del Alamo, Jes{\'u}s A and others},
  journal={Nature communications},
  volume={11},
  number={1},
  pages={3134},
  year={2020},
  publisher={Nature Publishing Group UK London}
}

@article{klyukin2018,
    author = {Klyukin, Konstantin and Rosso, Kevin M. and Alexandrov, Vitaly},
    title = {Iron Dissolution from Goethite (aplpha-FeOOH) Surfaces in Water by Ab Initio Enhanced Free-Energy Simulations},
    journal = {The Journal of Physical Chemistry C},
    volume = {122},
    number = {28},
    pages = {16086-16091},
    year = {2018},
    month = {07},
    issn = {1932-7447},
    doi = {10.1021/acs.jpcc.8b03743},
    url = {https://doi.org/10.1021/acs.jpcc.8b03743},
    eprint = {https://pubs.acs.org/jpccck/article-pdf/122/28/16086/15611034/jp8b03743.pdf},
}

@incollection{reticcioli2019,
  author    = {Reticcioli, Michele and Diebold, Ulrike and Kresse, Georg and Franchini, Cesare},
  title     = {Small polarons in transition metal oxides},
  booktitle = {Handbook of Materials Modeling},
  publisher = {Springer},
  year      = {2019}
}

@article{maxisch2006,
  author  = {Maxisch, Thomas and Zhou, Fei and Ceder, Gerbrand},
  title   = {Ab initio study of the migration of small polarons in olivine {Li}$_x${FePO}$_4$},
  journal = {Phys. Rev. B},
  volume  = {73},
  pages   = {104301},
  year    = {2006}
}

@article{wu2026tio2,
  author  = {Wu, Tongwei and Guo, Xiaoxi and Zhang, Guangjie and Zhang, Yanning and Song, Li and Liu, Zheng and Zhang, Hui and Shi, Shucheng and Liu, Limin and Melander, Marko M. and Honkala, Karoliina},
  title   = {Potential-dependent polaron formation activates {TiO}$_2$ for the hydrogen evolution reaction},
  journal = {Nat. Commun.},
  volume  = {17},
  pages   = {2104},
  year    = {2026}
}

@article{chretien2015,
  author  = {Chr\'etien, Steeve and Metiu, Horia},
  title   = {Hydrogen dissociative adsorption on lanthana: {Polaron} formation and the role of acid--base interactions},
  journal = {J. Phys. Chem. C},
  volume  = {119},
  pages   = {19876},
  year    = {2015}
}

@article{nagatsuka2022,
  author  = {Nagatsuka, Nozomi and Kato, Kensuke and Wilde, Markus and Fukutani, Katsuyuki},
  title   = {Absence of midgap states due to excess electrons donated by adsorbed hydrogen on the anatase {TiO}$_2$(101) surface},
  journal = {Phys. Rev. B},
  volume  = {105},
  pages   = {045424},
  year    = {2022}
}

@article{ziwritsch2016,
  author  = {Ziwritsch, Maximilian and M\"uller, Sonja and Hempel, Hannes and Unold, Thomas and Abdi, Fatwa F. and van de Krol, Roel and Friedrich, Dennis and Eichberger, Rainer},
  title   = {Direct time-resolved observation of carrier trapping and polaron conductivity in {BiVO}$_4$},
  journal = {ACS Energy Lett.},
  volume  = {1},
  pages   = {888},
  year    = {2016}
}

@article{wu2023bivo4,
  author  = {Wu, Hao and Zhang, Lei and Qu, Songying and Du, Aijun and Tang, Junwang and Ng, Yun Hau},
  title   = {Polaron-mediated transport in {BiVO}$_4$ photoanodes for solar water oxidation},
  journal = {ACS Energy Lett.},
  volume  = {8},
  pages   = {2177},
  year    = {2023}
}

@article{setvin2014,
  author  = {Setvin, Martin and Franchini, Cesare and Hao, Xianfeng and Schmid, Michael and Janotti, Anderson and Kaltak, Merzuk and Van de Walle, Chris G. and Kresse, Georg and Diebold, Ulrike},
  title   = {Direct view at excess electrons in {TiO}$_2$ rutile and anatase},
  journal = {Phys. Rev. Lett.},
  volume  = {113},
  pages   = {086402},
  year    = {2014}
}

@article{deskins2007,
  author  = {Deskins, N. Aaron and Dupuis, Michel},
  title   = {Electron transport via polaron hopping in bulk {TiO}$_2$: A density functional theory characterization},
  journal = {Phys. Rev. B},
  volume  = {75},
  pages   = {195212},
  year    = {2007}
}

@article{morgan2010,
  author  = {Morgan, Benjamin J. and Watson, Graeme W.},
  title   = {Intrinsic $n$-type defect formation in {TiO}$_2$: A comparison of rutile and anatase from {GGA}$+U$ calculations},
  journal = {J. Phys. Chem. C},
  volume  = {114},
  pages   = {2321},
  year    = {2010}
}

@article{elmaslmane2018,
  author  = {Elmaslmane, Amina R. and Watkins, Michael B. and McKenna, Keith P.},
  title   = {First-principles modeling of polaron formation in {TiO}$_2$ polymorphs},
  journal = {J. Chem. Theory Comput.},
  volume  = {14},
  pages   = {3740},
  year    = {2018}
}

@article{huang2017,
  author  = {Huang, Shibo and Verma, Prakash and Truhlar, Donald G.},
  title   = {Localizing holes as polarons and predicting band gaps, defect levels, and delithiation energies of solid-state materials with a local exchange-correlation functional},
  journal = {J. Phys. Chem. C},
  volume  = {121},
  pages   = {23955},
  year    = {2017}
}

@article{adelstein2014,
  author  = {Adelstein, Nicholas and Neaton, Jeffrey B. and Asta, Mark and De Jonghe, Lutgard C.},
  title   = {Density functional theory based calculation of small-polaron mobility in hematite},
  journal = {Phys. Rev. B},
  volume  = {89},
  pages   = {245115},
  year    = {2014}
}

@article{deng2023chgnet,
  author  = {Deng, Bowen and Zhong, Peichen and Jun, KyuJung and Riebesell, Janosh and Han, Kevin and Bartel, Christopher J. and Ceder, Gerbrand},
  title   = {{CHGNet} as a pretrained universal neural network potential for charge-informed atomistic modelling},
  journal = {Nat. Mach. Intell.},
  volume  = {5},
  pages   = {1031},
  year    = {2023}
}

@article{batatia2023mace,
    author = {Batatia, Ilyes and Benner, Philipp and Chiang, Yuan and Elena, Alin M. and Kovács, Dávid P. and Riebesell, Janosh and Advincula, Xavier R. and Asta, Mark and Avaylon, Matthew and Baldwin, William J. and Berger, Fabian and Bernstein, Noam and Bhowmik, Arghya and Bigi, Filippo and Blau, Samuel M. and Cărare, Vlad and Ceriotti, Michele and Chong, Sanggyu and Darby, James P. and De, Sandip and Della Pia, Flaviano and Deringer, Volker L. and Elijošius, Rokas and El-Machachi, Zakariya and Fako, Edvin and Falcioni, Fabio and Ferrari, Andrea C. and Gardner, John L. A. and Gawkowski, Mikołaj J. and Genreith-Schriever, Annalena and George, Janine and Goodall, Rhys E. A. and Grandel, Jonas and Grey, Clare P. and Grigorev, Petr and Han, Shuang and Handley, Will and Heenen, Hendrik H. and Hermansson, Kersti and Ho, Cheuk Hin and Hofmann, Stephan and Holm, Christian and Jaafar, Jad and Jakob, Konstantin S. and Jung, Hyunwook and Kapil, Venkat and Kaplan, Aaron D. and Karimitari, Nima and Kermode, James R. and Kourtis, Panagiotis and Kroupa, Namu and Kullgren, Jolla and Kuner, Matthew C. and Kuryla, Domantas and Liepuoniute, Guoda and Lin, Chen and Margraf, Johannes T. and Magdău, Ioan-Bogdan and Michaelides, Angelos and Moore, J. Harry and Naik, Aakash A. and Niblett, Samuel P. and Norwood, Sam Walton and O’Neill, Niamh and Ortner, Christoph and Persson, Kristin A. and Reuter, Karsten and Rosen, Andrew S. and Rosset, Louise A. M. and Schaaf, Lars L. and Schran, Christoph and Shi, Benjamin X. and Sivonxay, Eric and Stenczel, Tamás K. and Sutton, Christopher and Svahn, Viktor and Swinburne, Thomas D. and Tilly, Jules and van der Oord, Cas and Vargas, Santiago and Varga-Umbrich, Eszter and Vegge, Tejs and Vondrák, Martin and Wang, Yangshuai and Witt, William C. and Wolf, Thomas and Zills, Fabian and Csányi, Gábor},
    title = {A foundation model for atomistic materials chemistry},
    journal = {The Journal of Chemical Physics},
    volume = {163},
    number = {18},
    pages = {184110},
    year = {2025},
    month = {11},
    issn = {0021-9606},
    doi = {10.1063/5.0297006},
    url = {https://doi.org/10.1063/5.0297006},
    eprint = {https://pubs.aip.org/aip/jcp/article-pdf/doi/10.1063/5.0297006/20801246/184110_1_5.0297006.pdf},
}

@article{wood2025uma,
  author        = {Wood, Brandon M. and Dzamba, Misko and Fu, Xiang and Gao, Meng and Shuaibi, Muhammed and Barroso-Luque, Luis and Abdelmaqsoud, Kareem and Gharakhanyan, Vahe and Kitchin, John R. and Levine, Daniel S. and Michel, Kyle and Sriram, Anuroop and Cohen, Taco and Das, Abhishek and Rizvi, Ammar and Sahoo, Sushree Jagriti and Ulissi, Zachary W. and Zitnick, C. Lawrence},
  title={UMA: A family of universal models for atoms},
  journal={Advances in Neural Information Processing Systems},
  volume={38},
  pages={129391--129427},
  year={2026}
}

@article{zhou2026matris,
  author        = {Zhou, Yuanchang and Hu, Siyu and Zhang, Xiangyu and Wang, Hongyu and Tan, Guangming and Jia, Weile},
  title         = {{MatRIS}: Toward Reliable and Efficient Pretrained Machine Learning Interatomic Potentials},
  journal={arXiv preprint arXiv:2603.02002},
  year          = {2026},
}

@article{barroso2024omat24,
  title={The Open Materials 2024 (OMat24) inorganic materials dataset and models},
  author={Barros-Luque, Luis and Shuaibi, Muhammed and Fu, Xiang and Wood, Brandon M and Dzamba, Misko and Gao, Meng and Rizvi, Ammar and Uyttendaele, Matt and Zitnick, C Lawrence and Ulissi, Zachary W},
  journal={Nature Computational Science},
  pages={1--11},
  year={2026},
  publisher={Nature Publishing Group US New York}
}

@article{jain2013,
  author  = {Jain, Anubhav and Ong, Shyue Ping and Hautier, Geoffroy and Chen, Wei and Richards, William Davidson and Dacek, Stephen and Cholia, Shreyas and Gunter, Dan and Skinner, David and Ceder, Gerbrand and Persson, Kristin A.},
  title   = {Commentary: The {Materials Project}: A materials genome approach to accelerating materials innovation},
  journal = {APL Mater.},
  volume  = {1},
  pages   = {011002},
  year    = {2013}
}

@article{kresse1996,
  author  = {Kresse, Georg and Furthm\"uller, J\"urgen},
  title   = {Efficient iterative schemes for ab initio total-energy calculations using a plane-wave basis set},
  journal = {Phys. Rev. B},
  volume  = {54},
  pages   = {11169},
  year    = {1996}
}

@article{henkelman2000,
  author  = {Henkelman, Graeme and Uberuaga, Blas P. and J\'onsson, Hannes},
  title   = {A climbing image nudged elastic band method for finding saddle points and minimum energy paths},
  journal = {J. Chem. Phys.},
  volume  = {113},
  pages   = {9901},
  year    = {2000}
}

@article{schmidt2024alexandria,
  author  = {Schmidt, Jonathan and Cerqueira, Tiago F. T. and Romero, Aldo H. and Loew, Antoine and J\"ager, Fabian and Wang, Hai-Chen and Botti, Silvana and Marques, Miguel A. L.},
  title   = {Improving machine-learning models in materials science through large datasets},
  journal = {Mater. Today Phys.},
  volume  = {48},
  pages   = {101560},
  year    = {2024}
}

@article{mpaloe2025,
  author  = {Kuner, Matthew C. and Kaplan, Aaron D. and Persson, Kristin A. and Asta, Mark and Chrzan, Daryl C.},
  title   = {{MP-ALOE}: an r$^2${SCAN} dataset for universal machine learning interatomic potentials},
  journal = {npj Comput. Mater.},
  volume  = {11},
  pages   = {352},
  year    = {2025}
}

@article{Korostelev2026,
  author  = {Korostelev, Vladislav and \v{Z}guns, Pjotrs and Klyukin, Konstantin},
  title   = {Hydrogen in brownmillerite perovskites: First-principles insights into energetics and induced electronic-magnetic changes},
  journal = {Phys. Rev. Materials},
  volume  = {10},
  pages   = {044605},
  year    = {2026}
}

@article{zguns2026,
  author  = {\v{Z}guns, Pjotrs and Yildiz, Bilge},
  title   = {Effect of protons on polaron mobility in transition metal oxides},
  journal = {Phys. Rev. Materials},
  volume  = {10},
  pages   = {035402},
  year    = {2026}
}

@article{castleton2019,
  author  = {Castleton, Christopher W. M. and Lee, Amy and Kullgren, Jolla},
  title   = {Benchmarking density functional theory functionals for polarons in oxides: properties of {CeO}$_2$},
  journal = {J. Phys. Chem. C},
  volume  = {123},
  pages   = {5164},
  year    = {2019}
}

@article{kaplan2025matpes,
  author        = {Kaplan, Aaron D. and Liu, Runze and Qi, Ji and Ko, Tsz Wai and Deng, Bowen and Riebesell, Janosh and Ceder, Gerbrand and Persson, Kristin A. and Ong, Shyue Ping},
  title={A foundational potential energy surface dataset for materials},
  journal={arXiv preprint arXiv:2503.04070},
  year={2025}
}

@article{yalcin2026polflow,
  author  = {Yalcin, Firat and Verdi, Carla and Birschitzky, Viktor C. and Meier, Matthias and Wolloch, Michael and Reticcioli, Michele},
  title   = {Automated modeling of polarons: defects and reactivity on {TiO}$_2$(110) surfaces},
  journal = {npj Comput. Mater.},
  volume  = {12},
  pages   = {254},
  year    = {2026}
}

@article{birschitzky2024mlprediction,
  author  = {Birschitzky, Viktor C. and Ellinger, Florian and Franchini, Cesare and Reticcioli, Michele},
  title   = {Machine learning-based prediction of polaron-vacancy patterns on the {TiO}$_2$(110) surface},
  journal = {npj Comput. Mater.},
  volume  = {10},
  pages   = {89},
  year    = {2024}
}

@article{birschitzky2022mlconfig,
  author  = {Birschitzky, Viktor C. and Ellinger, Florian and Diebold, Ulrike and Reticcioli, Michele and Franchini, Cesare},
  title   = {Machine learning for exploring small polaron configurational space},
  journal = {npj Comput. Mater.},
  volume  = {8},
  pages   = {125},
  year    = {2022}
}

@article{leoni2026surfacepolarons,
  title={Machine-learned dynamics of surface polarons at reduced oxide surfaces},
  author={Leoni, Luca and Franchini, Cesare},
  journal={The Journal of Physical Chemistry Letters},
  volume={17},
  number={32},
  pages={9237--9245},
  year={2026},
  publisher={ACS Publications}
}

@article{birschitzky2025leopold,
  author  = {Birschitzky, Viktor C. and Leoni, Luca and Reticcioli, Michele and Franchini, Cesare},
  title   = {Machine Learning Small Polaron Dynamics},
  journal = {Phys. Rev. Lett.},
  volume  = {134},
  pages   = {216301},
  year    = {2025}
}

@article{dai2025polaronsreview,
  author        = {Dai, Zhenbang and Lafuente-Bartolome, Jon and Giustino, Feliciano},
  title={Polarons from first principles},
  journal={arXiv preprint arXiv:2512.06176},
  year={2026}
}

@article{zhong2025mnrich,
  author  = {Zhong, Peichen and Deng, Bowen and Anand, Shashwat and Mishra, Tara and Ceder, Gerbrand},
  title   = {Modeling phase transformations in {Mn}-rich disordered rocksalt cathodes with machine-learning interatomic potentials},
  journal = {Phys. Rev. Mater.},
  volume  = {9},
  pages   = {105404},
  year    = {2025},
  doi     = {10.1103/mk2d-tjyj}
}

\end{document}


\section{Methods}

\emph{Reference DFT$+U$ calculations.} All reference calculations were performed with the Vienna Ab initio Simulation Package (VASP), version 6.3.0~\cite{kresse1996}. We used the projector augmented-wave (PAW) method with the PBE exchange-correlation functional, and all calculations were spin-polarized. A Hubbard $U$ correction was applied to the transition-metal $d$ states using the values adopted by the Materials Project~\cite{jain2013}. These are 3.32~eV for Co, 3.25~eV for V, 3.70~eV for Cr, 3.90~eV for Mn, 4.38~eV for Mo, 5.30~eV for Fe, and 6.20~eV for both Ni and W. No $U$ was applied to Ti and Ce, except in the TiO$_2$ sweep of Fig.~1, where $U$(Ti) was varied from 0 to 5~eV. The plane-wave cutoff was 520~eV. Structures were relaxed until the residual forces fell below 0.02~eV/\AA{} (\texttt{EDIFFG}~$=-0.02$), and each electronic self-consistency loop was converged to $1\times10^{-6}$~eV (\texttt{EDIFF}). Because of the large supercell volumes, a $2\times2\times2$ $\Gamma$-centered $k$-point mesh was used for all systems. Polaron-transfer barriers were obtained from converged climbing-image nudged elastic band (NEB) calculations~\cite{henkelman2000}.

\emph{Materials.}Our research employs a test set that spans various structure types and magnetic configurations. The spin configuration of each supercell is given after its composition. Rutile TiO$_2$ has 54 Ti, 108 O, and 1 H atoms (paramagnetic). The brownmillerites Sr$_2$Fe$_2$O$_5$ and Sr$_2$Co$_2$O$_5$ each have 32 transition-metal atoms, 32 Sr, and 80 O (AFM-G). Spinel Co$_3$O$_4$ has 24 Co, 32 O, and 1 H (AFM). Rock-salt NiO has 108 Ni, 108 O, and 1 H (FM). Layered V$_2$O$_5$ has 48 V, 120 O, and 1 H (paramagnetic). Olivine LiFePO$_4$ has 64 Fe, 64 Li, 64 P, and 256 O (AFM), and FePO$_4$ has 16 Fe, 16 P, and 64 O (AFM). CrO$_2$ has 32 Cr, 64 O, and 1 H (AFM-G), and corundum Cr$_2$O$_3$ has 48 Cr, 72 O, and 1 H (FM). Fig.~5 includes six additional hosts. MnO$_2$ has 32 Mn, 64 O, and 1 H (AFM). Fe$_2$O$_3$ has 128 Fe, 192 O, and 1 H (FM). Fe$_3$O$_4$ has 192 Fe, 256 O, and 1 H (FM). MoO$_3$ has 36 Mo, 108 O, and 1 H (paramagnetic). WO$_3$ has 72 W, 216 O, and 1 H (paramagnetic). CeO$_2$ has 32 Ce, 64 O, and 1 H (paramagnetic).

\emph{Hydrogen as electron donor.} Universal machine-learning interatomic potentials (uMLIPs) have no explicit electronic or charge degrees of freedom, so an excess electron cannot be added to the system directly. The open-source models used here accept only atomic positions and chemical species as input. We therefore introduce the electron chemically by doping with one interstitial H atom in each supercell (H$\to$H$^+ + e^-$). The donated electron then forms a polaron on a nearby transition-metal site. This approach was used for all hosts. For V$_2$O$_5$ we also tested two alternative electron donors, a Li interstitial (Li$\to$Li$^+ + e^-$) and an oxygen vacancy (Fig.~5). Transition-metal oxides can adopt several magnetic orderings (FM and different AFM types), and these give slightly different bond lengths and defect potential-energy surfaces. The uMLIPs are blind to spin. CHGNet and MatRIS predict only scalar magnetic-moment magnitudes.

\emph{$Z$-score evaluation.} For each transition-metal site, the first oxygen shell was defined by an element-specific distance cutoff, chosen from the gap in the radial distribution function. The cutoffs are 2.3~\AA{} for TiO$_2$, LiFePO$_4$, and FePO$_4$, 2.1~\AA{} for V$_2$O$_5$ and CrO$_2$, and 2.15~\AA{} for Cr$_2$O$_3$. For V$_2$O$_5$ the next neighbor appears only at 2.88~\AA. For Sr$_2$Fe$_2$O$_5$ and Sr$_2$Co$_2$O$_5$ the cutoff is 2.5~\AA, which separates the CN4 bonds (1.8--1.9~\AA) and CN6 bonds (2.0--2.1~\AA) from the next shell. In lattices with mixed coordination numbers, the $Z$-scores were standardized separately within each coordination sublattice. In lattices with a single coordination number, they were standardized over all transition-metal sites.

\emph{Machine-learning interatomic potentials.} We benchmarked five model checkpoints. CHGNet v0.3.0 has 412{,}525 parameters and was trained on MPtrj~\cite{deng2023chgnet}. MACE-MPA (\texttt{mace-mpa-0-medium}) was trained on MPtrj combined with Alexandria~\cite{batatia2023mace}, and MACE-OMAT (\texttt{mace-omat-0-medium}) on OMAT24. UMA (\texttt{uma-s-1p2}, \texttt{omat} task) was trained on OMAT24~\cite{wood2025uma}. MatRIS (\texttt{matris\_10m\_oam}) has 10{,}443{,}505 parameters and was trained on OMat24, Alexandria, and MPtrj jointly~\cite{zhou2026matris}. Where magnetic moments are reported, they come from CHGNet or MatRIS single-point calculations on the geometry in question.

\emph{Workflow.} The procedure has two stages: a DFT$+U$ stage that produces the reference geometries, and a uMLIP stage that starts from them.

\textbf{DFT$+U$ stage.} For most systems, one H atom was inserted in the interstitial position near an oxygen atom of the pristine supercell, and the structure was relaxed with DFT$+U$ without constraints. In some hosts H has several non-equivalent binding sites, and each site correspond to a different polaron geometry. This is the case for the brownmillerites Sr$_2$Fe$_2$O$_5$ and Sr$_2$Co$_2$O$_5$. For these systems, a direct relaxation sometimes moved H away from the intended site, so the desired local minimum could not be reached. In those cases we first relaxed the framework with H frozen at the chosen site, then released H and re-optimized the full structure. This gave a converged geometry for each H site and its associated polaron.

\textbf{uMLIP stage.} The converged DFT$+U$ geometries (see Data availability) were used as the starting structures for all uMLIP calculations, so that every model begins from the same localized polaron configuration. (1) The pristine structure was fully optimized, including cell and atomic positions, using a Fr\'echet cell filter. (2) The H-containing structure was then optimized with BFGS, and $Z_1/Z_2$ and $|\Delta\mu|$ were evaluated at every optimization step through a callback. (3) The final $Z_1/Z_2$ was compared with the DFT$+U$ reference. (4) To test the dependence on the starting geometry, we cross-initialized the models (DFT$\to$uMLIP and uMLIP$_A\to$uMLIP$_B$) and compared the energies of the localized and delocalized final states. (5) The potential-energy surface between the localized and delocalized endpoints was mapped by linear interpolation with 100 images, evaluating UMA and MatRIS single-point energies along the path.

\emph{Data availability.} All DFT+$U$ input and relaxed geometries (VASP POSCAR format), the tabulated data underlying every figure in the main text and this Supplement, and an example Google Colab notebook demonstrating the $Z$-score workflow on a uMLIP (structure upload, relaxation, and per-step $Z_1$/$Z_2$ tracking) are available at \url{https://doi.org/10.6084/m9.figshare.33758275}.

\begin{figure}
    \centering
    \includegraphics[width=\linewidth]{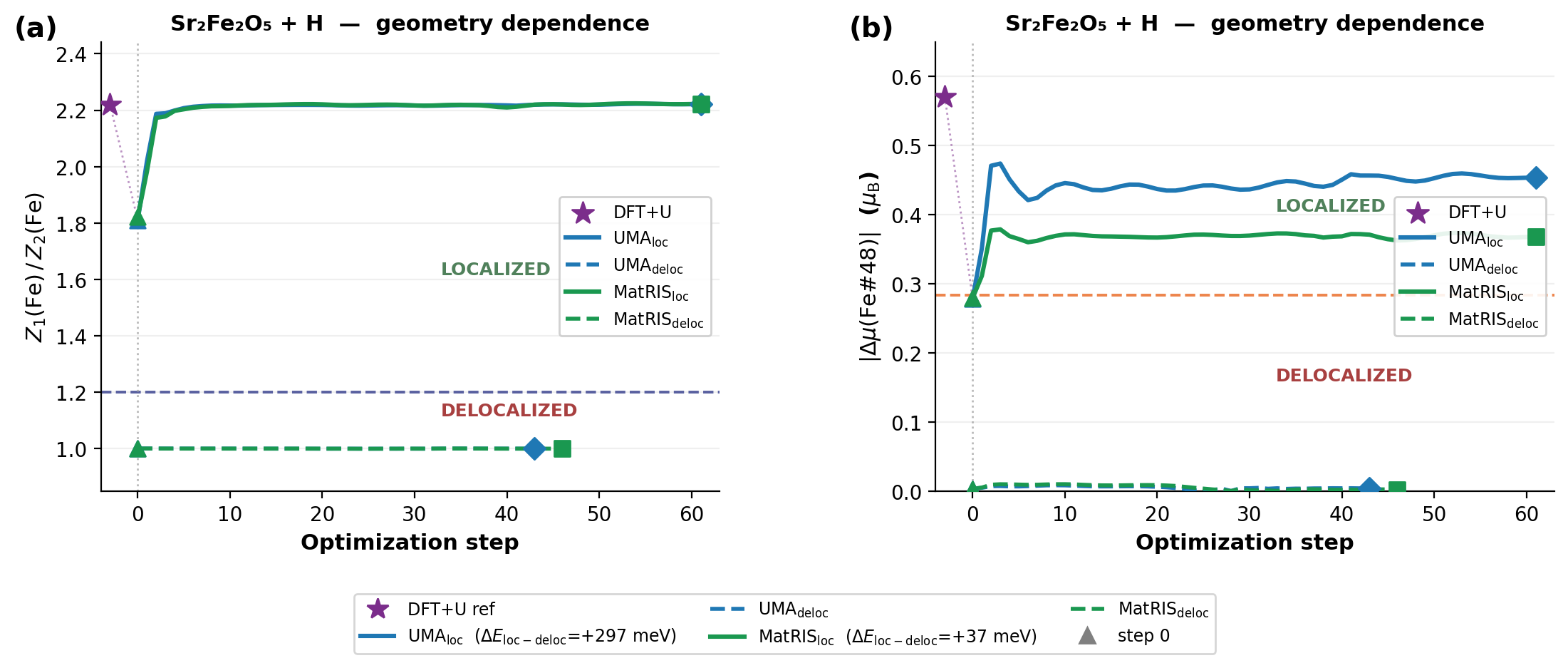}
    \caption{Geometry dependence of the relaxed polaron state in Sr$_2$Fe$_2$O$_5$+H for UMA and MatRIS. Optimizations are initialized from two distinct starting geometries -- one with a localized polaron (solid lines, triangle at step~0) and one with a delocalized polaron (dashed lines, triangle at step~0) -- and tracked to convergence (diamond: UMA endpoint; square: MatRIS endpoint). (a)~Polaron selectivity $Z_1(\mathrm{Fe})/Z_2(\mathrm{Fe})$ and (b)~maximum on-site magnetic-moment change $|\Delta\mu(\mathrm{Fe}\#48)|$ along each trajectory. Both UMA and MatRIS remain trapped in whichever basin -- localized or delocalized -- they are initialized in, confirming that both minima are genuine, coexisting solutions on their potential-energy surfaces rather than artifacts of a single relaxation path. Star: DFT$+U$ reference. The energy differences between the converged localized and delocalized states are $\Delta E_{\mathrm{loc-deloc}} = +297$~meV (UMA) and $+37$~meV (MatRIS).}
    \label{fig:sfo-geom-dependence}
\end{figure}

\begin{figure}
    \centering
    \includegraphics[width=\linewidth]{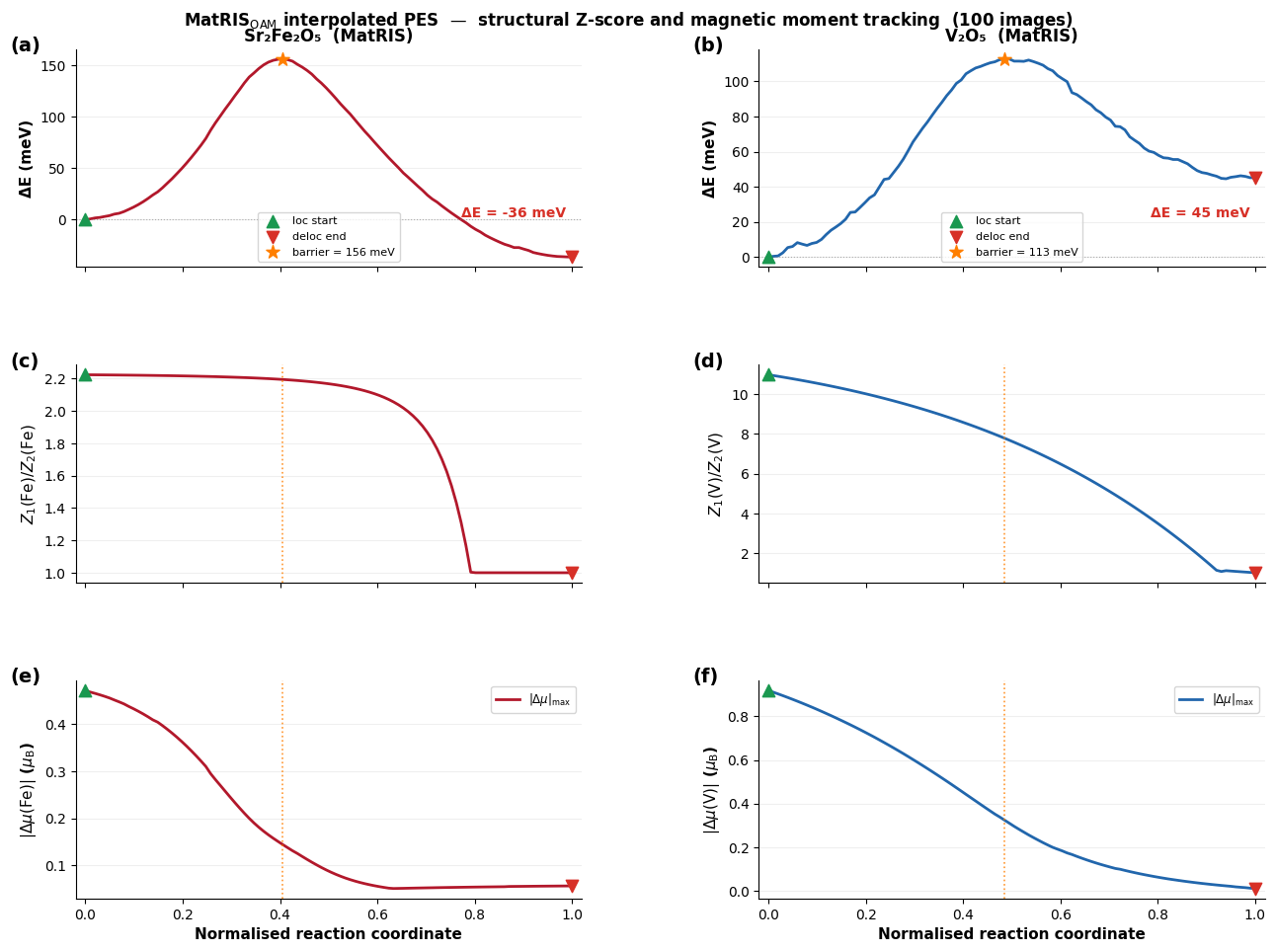}
    \caption{MatRIS$_\mathrm{OAM}$ interpolated PES between localized (DFT-initialized) and delocalized (CHGNet-initialized) endpoints for \ce{Sr2Fe2O5} (left) and \ce{V2O5} (right), 100 images, single-point energies. (a,b)~Relative energy $\Delta E$ along the normalized reaction coordinate; orange star marks the transition state. (c,d)~Polaron selectivity $Z_1/Z_2$ (CN-stratified for \ce{Sr2Fe2O5}; global for \ce{V2O5}). (e,f)~Maximum on-site magnetic-moment change $|\Delta\mu|_\mathrm{max}$, evaluated natively from MatRIS. For \ce{Sr2Fe2O5}, the delocalized state lies \SI{36}{\milli\electronvolt} below the localized one, behind a \SI{156}{\milli\electronvolt} barrier; for \ce{V2O5}, the localized state is instead the ground state, \SI{45}{\milli\electronvolt} lower, behind a \SI{113}{\milli\electronvolt} barrier. Dotted vertical line in (c)--(f): transition-state position from (a,b).}
    \label{fig:matris-pes}
\end{figure}

\begin{figure}
    \centering
    \includegraphics[width=\linewidth]{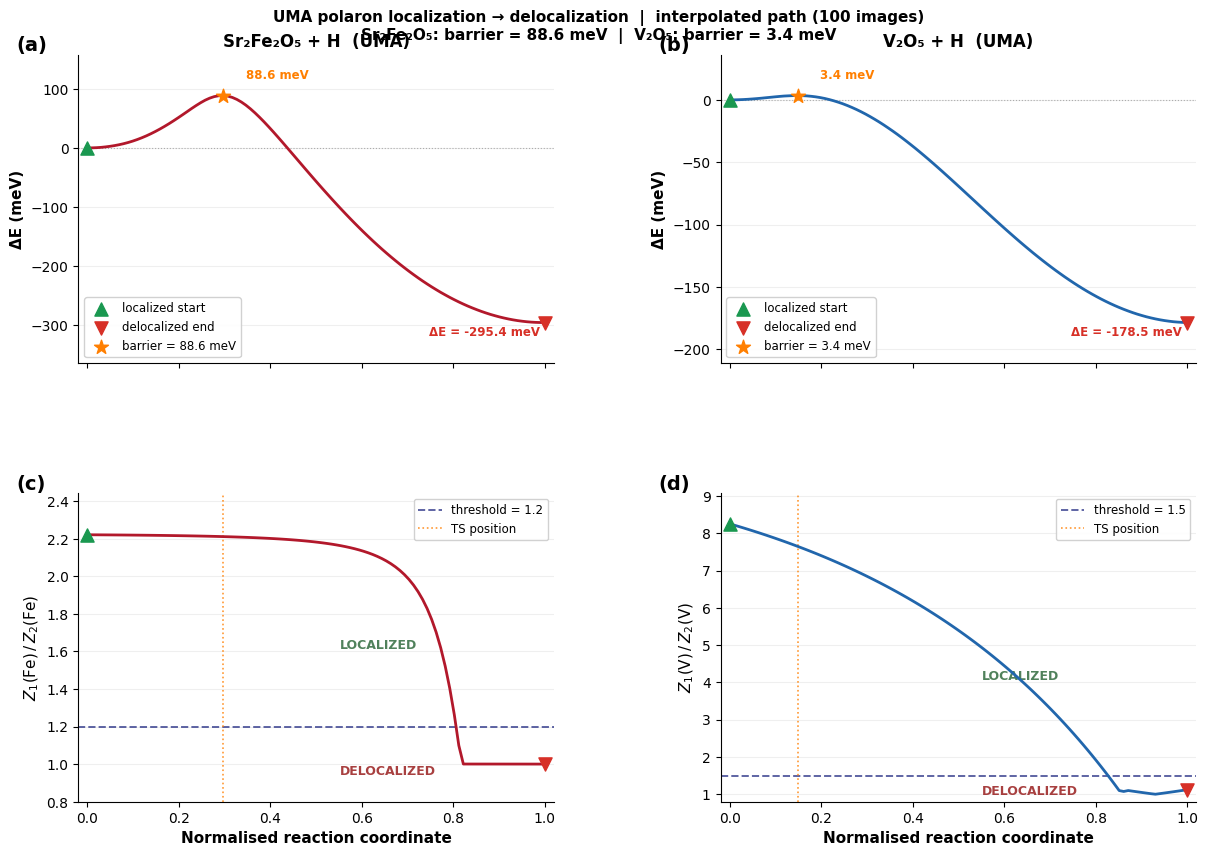}
    \caption{UMA interpolated PES between localized and delocalized polaron endpoints for Sr$_2$Fe$_2$O$_5$ (left) and V$_2$O$_5$ (right), 100 images. (a,b)~Relative energy $\Delta E$ along the normalized reaction coordinate; orange star marks the transition state. (c,d)~Polaron selectivity $Z_1/Z_2$, with dashed threshold separating localized and delocalized regimes. For both hosts the localized state lies above the delocalized minimum: $\Delta E = \SI{295.4}{\milli\electronvolt}$ with an \SI{88.6}{\milli\electronvolt} barrier for Sr$_2$Fe$_2$O$_5$, and $\Delta E = \SI{178.5}{\milli\electronvolt}$ with a \SI{3.4}{\milli\electronvolt} barrier for V$_2$O$_5$. Dotted vertical line in (c,d): transition-state position from (a,b).}
    \label{fig:uma-pes}
\end{figure}

\begin{figure}
    \centering
    \includegraphics[width=\linewidth]{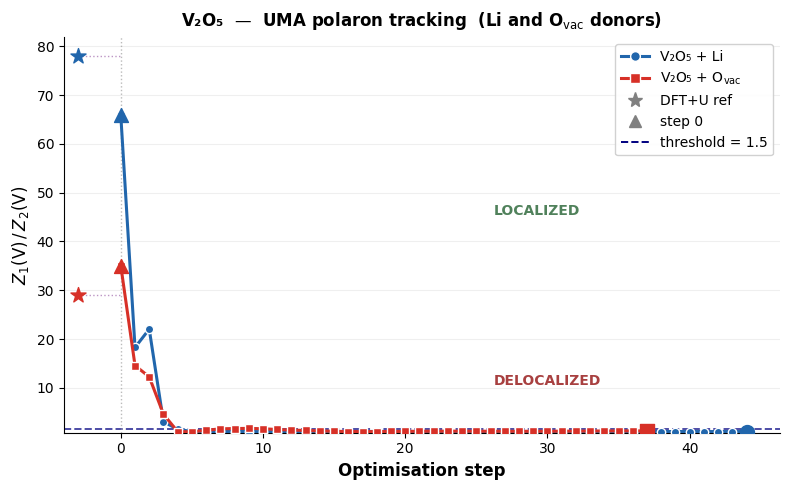}
    \caption{UMA polaron tracking in V$_2$O$_5$ for Li interstitial and O-vacancy electron donors. Both relaxations are initialized from a localized DFT$+U$ polaron geometry (triangle: step~0) and rapidly collapse to $Z_1/Z_2 \approx 1$ within $\sim$5 optimization steps, falling below the localization threshold (dashed line). Unlike the H-doped case, neither Li interstitials nor O vacancies sustain a localized polaron minimum on the UMA potential-energy surface for V$_2$O$_5$. Star: DFT$+U$ reference for each donor.}
    \label{fig:v2o5-li-ovac}
\end{figure}

\begin{figure}
    \centering
    \includegraphics[width=\linewidth]{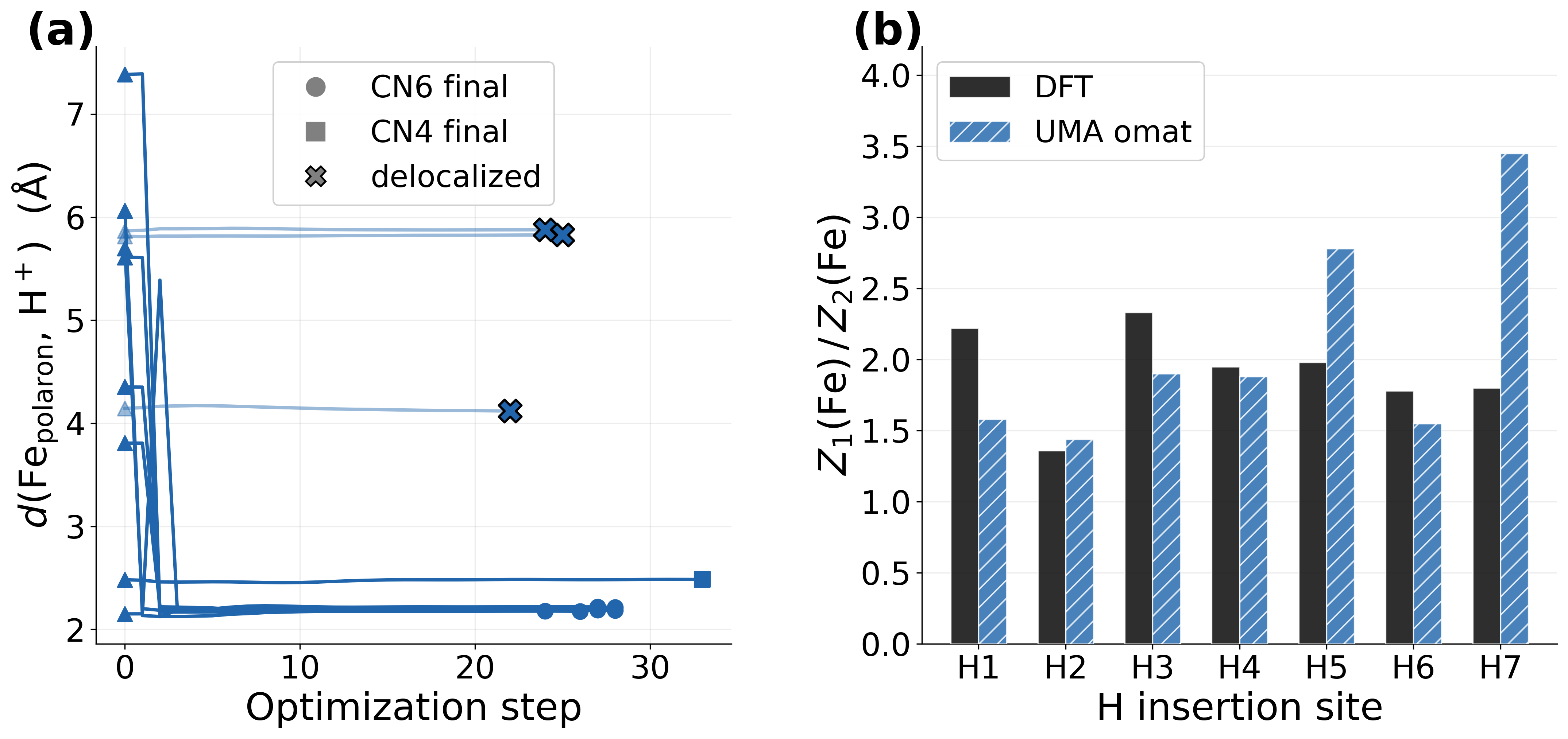}
    \caption{Polaron site selectivity in Sr$_2$Fe$_2$O$_5$ (UMA), analogous to the Sr$_2$Co$_2$O$_5$ analysis in the main text. (a)~Distance between the polaronic Fe site and the fixed proton (H$_1$) during relaxations initialized with the distortion on different Fe sites. Unlike Sr$_2$Co$_2$O$_5$, not all trajectories converge to a localized polaron: most collapse to one preferred site at $\sim$2.15--2.5~\AA{} (circles: CN6 final; squares: CN4 final), while a subset instead delocalize (crosses). (b)~$Z_1/Z_2$ for proton sites H$_1$--H$_7$, DFT (black) vs UMA (blue).}
    \label{fig:sfo-site-selectivity}
\end{figure}

\begin{figure}
    \centering
    \includegraphics[width=\linewidth]{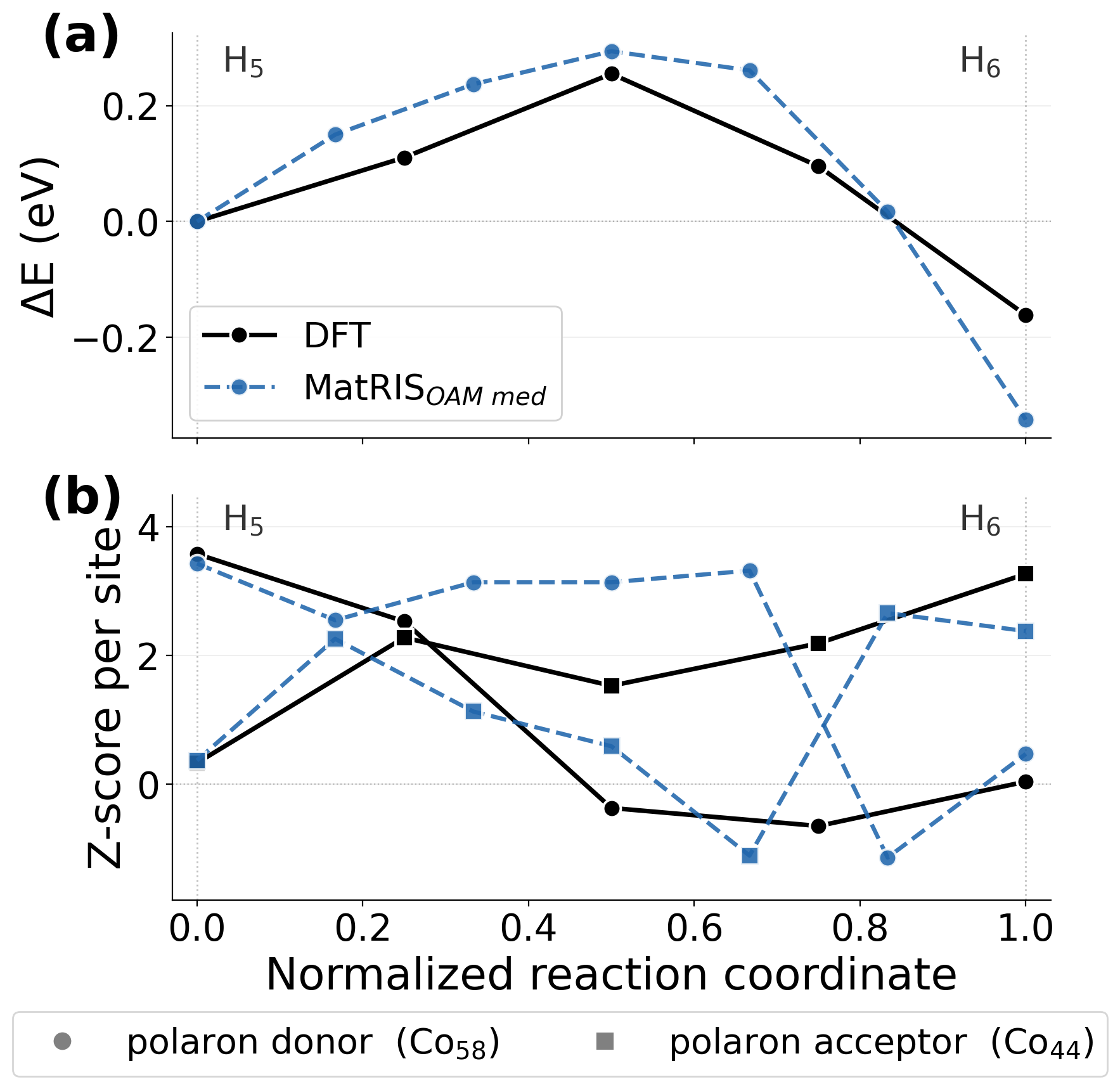}
    \caption{H$_5\to$H$_6$ polaron transfer in Sr$_2$Co$_2$O$_5$, MatRIS$_\mathrm{OAM\ med}$ counterpart to the UMA result in the main text. (a)~Energy profile: DFT$+U$ CI-NEB (black) vs MatRIS (blue dashed). (b)~$Z$-scores of donor (Co$_{58}$, circles) and acceptor (Co$_{44}$, squares) sites along the normalized reaction coordinate. MatRIS reproduces the barrier shape and the donor--acceptor $Z$-score handoff seen with UMA, tracking the DFT$+U$ profile closely near the transition state.}
    \label{fig:sco-neb-matris}
\end{figure}

\clearpage

\bibliography{references}